\documentclass{pprai}

\usepackage[utf8]{inputenc}
\usepackage[T1]{fontenc}
\usepackage{graphicx}
\usepackage{amsthm}
\usepackage{txfonts}
\usepackage{url}
\usepackage{algpseudocode}
\usepackage{tabularx}
\usepackage{booktabs}
\usepackage{float}

\title{Ransomware Detection Using Machine Learning in the Linux Kernel}
\headtitle{Ransomware Detection Using Machine Learning in the Linux Kernel}

\author{%
    Adrian Brodzik$^{1[0009-0009-5980-0620]}$,\\
    Tomasz Malec-Kruszy{\'n}ski$^{1[0009-0009-9555-8560]}$,\\
    Wojciech Niewolski$^{1[0000-0001-6779-9038]}$,\\
    Miko{\l}aj Tkaczyk$^{1[0009-0007-5462-6438]}$,\\
    Krzysztof Bocianiak$^{1}$,\\
    Sok-Yen Loui$^{2}$\\
}

\headauthor{A. Brodzik,  T. Kruszy{\'n}ski, W. Niewolski, M. Tkaczyk, K. Bocianiak, S. Loui}

\affiliation{%
    $^1$Orange Polska, Aleje Jerozolimskie 160, 02-326 Warsaw, Poland\\
    $^2$Orange, 44 Avenue de la R{\'e}publique, 92326 Ch{\^a}tillon, France\\
    firstname.lastname@orange.com
}


\begin{document}

\maketitle

\begin{abstract}
Linux-based cloud environments have become lucrative targets for ransomware attacks, employing various encryption schemes at unprecedented speeds. Addressing the urgency for real-time ransomware protection, we propose leveraging the extended Berkeley Packet Filter (eBPF) to collect system call information regarding active processes and infer about the data directly at the kernel level. In this study, we implement two Machine Learning (ML) models in eBPF - a decision tree and a multilayer perceptron. Benchmarking latency and accuracy against their user space counterparts, our findings underscore the efficacy of this approach.
\end{abstract}

\section{Introduction}

In recent times, there has been a notable surge in global cyberattacks, with ransomware emerging as a prominent threat. This malicious software encrypts user data and demands a ransom for its recovery. Modern variants are also capable of data theft and self-propagation. Globally, over 72\% of all organizations claim to have been affected by ransomware in 2023 - this number is not expected to decrease \cite{ibm2023cost}. In order to defend against such attacks, a low-level, efficient and extremely low latency tool has to be used. The extended Berkeley Packet Filter (eBPF) fits those criteria perfectly, as it allows developers to run programs within the operating system kernel without having to modify the kernel source code.

The eBPF programs can be written in a constrained C-like non-Turing-complete programming language and compiled to eBPF bytecode using the Clang/LLVM toolchain and libbpf library. The bytecode is dynamically loaded into the kernel's virtual machine and checked in terms of execution safety by the eBPF verifier. This simplifies the development process, as well as ensures kernel stability after deployment. Although eBPF programs are stateless, special data structures, called BPF maps, can be created, allowing data storage and memory sharing. This enables communication between eBPF programs and user space applications. The fundamental purpose of eBPF is system observability. Different events can be captured using kernel probes, user probes, and other tracepoints. Apart from that, the Linux Security Module (LSM) framework introduces various security checks which can be hooked. These can be used to block specific user and process operations, allowing the creation of dynamic security policies. As such, combining eBPF with Machine Learning (ML) algorithms may offer enhanced detection capabilities. In this paper, we propose an innovative solution based on ML models embedded in eBPF programs for the purpose of real-time ransomware detection.

\section{Related Work}

Prior research has extensively explored the domain of ransomware detection, especially for Windows-based operating systems \cite{SurveyRansomwareIJCSIS2018, urooj2021ransomware}. The methods used range from static, dynamic, and hybrid analysis techniques, including ML models \cite{alhawi2018leveraging, moussaileb2020ransomware}. Deceptive techniques like deploying Honeypots and Honeyfiles have also proved to be effective early detection mechanisms \cite{el2018intrusion}. Currently, the use of ML systems is considered to be the most effective (especially against zero-day attacks). Different features can be utilized, such as those extracted from network traffic \cite{almashhadani2019multi}.

Many ML-based solutions are typically implemented in high-level programming languages and consequently reside in the user space. This placement introduces substantial latencies in the decision-making process. Given the speed of ransomware cryptographic algorithms, even seemingly inconsequential delays can lead to a higher volume of encrypted files. Optimization of the time between data acquisition and the final decision is critical. The use of ML in eBPF can be a significant step forward in this regard. Currently only a couple research papers exist that touch on this subject; many of which use eBPF for process observability and ML inference in the user space \cite{higuchi2023real}. The closest implementation involves the use of eBPF in an Intrusion Detection System (IDS), demonstrating enhanced performance of decision tree models in the kernel space compared to their user space versions \cite{bachl2021flow}. This improvement is also observed for Random Forest, Support Vector Machine (SVM), and TwinSVM written in eBPF \cite{anand2023high}. Both of these studies rely on network communication and focus on general intrusion detection. Our solution, on the other hand, explores the ransomware-specific threat model and utilizes features extracted from process activity.

The lack of research papers may serve both as a challenge and an opportunity for future research work in the rapidly expanding domain of ransomware attacks. Due to that fact, addressing the limitation of existing literature may unlock new insights and bring meaningful advancement within the field. Effectively optimizing decision-making through ML could mitigate the impact of future attacks, making them less detrimental and more challenging for potential adversaries to conduct.

\section{Ransomware Detection in eBPF}

More and more ransomware attacks are targeting Linux-based cloud environments. Therefore, it is essential to develop appropriate security measures to protect data from encryption and exfiltration. Instead of analyzing executable files, the focus is shifted on the behavioral aspects of ransomware. We assume that an attacker has compromised a developer account, can execute arbitrary commands, but does not have root access. The ransomware attack does not have to be a single executable binary file – it can be multiple commands (or a script) utilizing benign system administrative tools. Ransomware can be multi-threaded and multi-process, and can implement various techniques and encryption schemes.

\subsection{Implementation}

\begin{figure}[h]
    \centering
    \includegraphics[width=\textwidth]{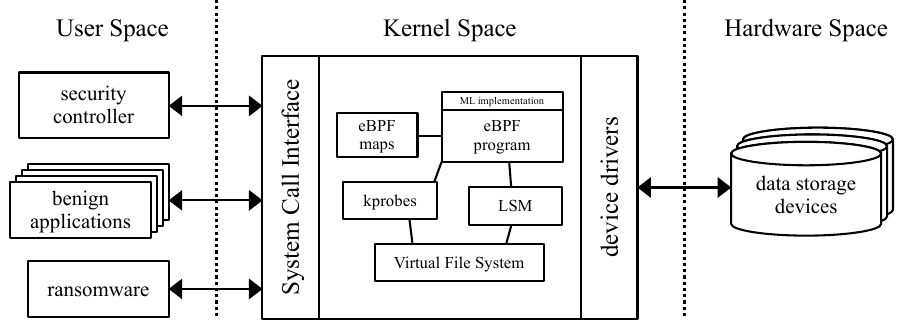}
    \caption{Real-time ransomware detection architecture overview.}
    \label{fig:arch1}
\end{figure}

Our ransomware detection application protects a set of directory paths, e.g. containerized application data volumes. It is based on several key components. First, eBPF is used to monitor and count selected system calls and other events triggered by different processes. To overcome the overwhelming stream of information the data is filtered by the directory path it pertains to or by process ID which has accessed a protected path in the past. These events were selected based on prior ransomware analysis efforts and include: \textit{file\_permission}, \textit{file\_open}, \textit{inode\_create}, \textit{inode\_unlink}, \textit{inode\_rmdir}, \textit{inode\_rename}, \textit{getdents64}, \textit{vfs\_read}, \textit{vfs\_write}. More information can be found in the LSM\footnote{\url{https://www.kernel.org/doc/html/v5.1/security/LSM.html}} and Syscall\footnote{\url{https://www.man7.org/linux/man-pages/man2/syscalls.2.html}} documentation. Additionally, Shannon entropy and Pearson's $\chi^2$ goodness-of-fit metric are calculated for write operations. Secondly, the data is relayed from the kernel space to user space applications via eBPF data structure maps, such as ring buffers. There, the events can be processed using ML algorithms to distinguish benign and malicious activity. Finally, the process or entire user account can be blocked, access to a particular path can be revoked via traditional Linux capabilities or the Linux Security Module (LSM) framework. This method may introduce significant delays, which jeopardizes file integrity, given the speed of modern ransomware encryption schemes. Instead of sending event data and analyzing it in the user space, we attempt to implement the entire threat detection process directly in eBPF – reducing latency without losing accuracy. An overview of the architecture is shown in Figure \ref{fig:arch1}.

For the purpose of this study, a decision tree and a multilayer perceptron were trained in a supervised matter to classify process events as either benign or malicious. The models were prepared using Scikit-Learn and PyTorch. Internally, a single decision tree structure is stored as several arrays, such as the feature used for splitting the node, the identifiers of the left and right child nodes, and the threshold value used to decide which child node to select next. The decision tree classifier was trained using default parameters with no depth or node count limitations. The final decision tree has 97 nodes. Similarly, PyTorch stores the parameters of each linear layer in their corresponding weight and bias tensors, which can be represented as 1D and 2D arrays. The neural network was trained by minimizing binary cross-entropy loss using Adam optimizer with learning rate $\gamma=10^{-4}$ for no more than 10000 epochs (until loss stopped decreasing). The final neural network has two linear layers ($12\rightarrow8$ and $8\rightarrow1$ features) with ReLU and sigmoid activation functions, totaling 113 parameters. All aforementioned models were retrained at least 20 times with different random number generator seeds to ensure consistency.

These models were chosen because of their simplicity and efficiency, while granting the ability to represent nonlinear data relations. Their functionality was replicated in C and eBPF, including small matrix multiplication and array-based tree traversal. Because eBPF is a restrictive programming environment, complex models like deep neural networks may be near impossible to implement. Decision trees and small neural networks, on the other hand, can be easily constructed and executed within the eBPF constraints, making them suitable for such use cases. Due to the limitations of eBPF, including the lack of floating-point operations, the models had to be adapted. All floating-point operations had to be converted to fixed-point arithmetic. Furthermore, the sigmoid function was not implemented, instead its input logits were used for thresholding. Model parameters were dynamically loaded using eBPF data structures or stored as hard-coded values in the read-only data section of the program.

\subsection{Methodology}

The aforementioned ransomware detection models were tested in a Proxmox virtual environment with two virtual machines. The first machine served as a single cluster node, running a database application that we wished to protect. The second machine acted as a benign traffic generator, sending HTTP requests to read, write, and delete random files of various sizes (1 KB - 1 GB), random names, and realistic file extensions (office documents, images, source code files). The contents were comprised of randomly generated Lorem Ipsum texts and sometimes compressed using LZMA. A dataset of 10 Linux ransomware samples was created and split into train and test sets, including Royal, Conti, Monti, HelloKitty, Kuiper, IceFire, RansomEXX, Buhti, BlackBasta, Hive, Cl0p, RedAlert. These samples were obtained from the MalwareBazaar database and represent several infamous Ransomware as a Service (RaaS) threat actors, which have started infiltrating Linux, Cloud, VMware ESXi, and similar environments, having previously targeted Windows-based systems. The chosen families have been analyzed in terms of historical significance. This way we achieve diversity in evasion, propagation, enumeration, and encryption techniques. The malicious programs were run independently one by one with and without client traffic, as well as multiple samples simultaneously. After each experiment the test environment was rolled back to the previous snapshot. The objective was to measure the processing time and macro $F_1$ classification score across the different models in user space Python, C, and kernel space eBPF.

\section{Experimental Results}

For each test case, we recorded raw event data, process system call statistics, processing time, and predictions of our models. The final results have been aggregated in Table \ref{tab:results}. Because the tests were conducted in a live environment, the sample size and dataset imbalance may vary (there are approximately twice as many malicious events). Overall, the eBPF implementations of the decision tree and multilayer perceptron yield a significant performance boost over their user space versions, without negatively affecting precision and recall. All models achieved macro $F_1>95\%$, accurately separating ransomware activity from this particular client network traffic simulator. As such, it would be beneficial to conduct more tests with varying traffic simulator algorithms and a less constrained environment, further masking ransomware activity, in the future.

\begin{table}[hb]
\centering
\caption{Decision tree and neural network experimental results.}
\label{tab:results}
\begin{tabular}{cccrrr}
\toprule
& & & \multicolumn{3}{c}{processing time (in nanoseconds)} \\
\cmidrule{4-6}
implementation & sample size & macro $F_1$ & mean & \begin{tabular}{@{\hspace{0.75em}}@{}c@{}}standard\\deviation\end{tabular} & median \\
\midrule
\multicolumn{6}{c}{\textbf{decision tree}} \\
\midrule
Python & 51601 & 0.997 & 167183 & 150271 & 150759 \\
C & 35400 & 0.998 & 818 & 1196 & 714 \\
eBPF & 36525 & 0.998 & 115 & 301 & 93 \\
\midrule
\multicolumn{6}{c}{\textbf{neural network}} \\
\midrule
Python & 32165 & 0.953 & 94996 & 103528 & 83910 \\
C & 35410 & 0.965 & 1062 & 1304 & 754 \\
eBPF & 37910 & 0.974 & 220 & 488 & 180 \\
\bottomrule
\end{tabular}
\end{table}

Based on the average processing time, the eBPF implementation of the decision tree is 7.1x faster than the C version and 1453x faster than the Python version. Similarly, the eBPF implementation of the neural network is 4.8x faster than the C version and 431x faster than the Python version. The reason for such differences is because Python is an interpreted language, while C and eBPF are compiled. Even though the eBPF and C versions are almost identical, sending data from the kernel space, analyzing it in the user space, and sending back the classification result has a significant impact on the total processing time, delaying ransomware detection.

\section{Conclusion}

This paper investigates the integration of two ML algorithms, decision tree and multilayer perceptron, in eBPF with the objective of enhancing early ransomware detection in Linux environments. The results demonstrate the viability of this solution. In the presence of extremely fast encryption algorithms used by modern ransomware, this solution can be used to minimize latencies and enable real-time ransomware detection. Our ongoing research efforts focus on developing dedicated ML solutions tailored for the identification of ransomware activity. By implementing these models inside the Linux kernel we aim to achieve better detection results in terms of both effectiveness and time efficiency. Building upon the successful application of the decision tree classifier in the Linux kernel, our future work involves implementing other supervised and unsupervised algorithms in eBPF, such as Isolation Forests, Random Forests, and Support Vector Machines.

\newpage
\bibliography{pprai}
\bibliographystyle{pprai}

\end{document}